\documentclass[epj]{svjour}
\usepackage{latexsym}
\usepackage{graphicx}
\usepackage{subfigure}
\expandafter\let\csname equation*\endcsname\relax
\expandafter\let\csname endequation*\endcsname\relax
\usepackage{amsmath}
\usepackage{amssymb}
\usepackage[square,sort&compress]{natbib}
\usepackage{array}
\newcommand{\eref}[1]{(\ref{#1})}
\newcommand{\scs}{\scriptscriptstyle}
\newcommand{\ds}{\displaystyle}
\begin{document}
\title{Persistent random walk with exclusion}
\subtitle{Persistent random walk with exclusion}
\author{Marta Galanti\inst{1,2,3} \and Duccio Fanelli\inst{1,3} \and Francesco Piazza\inst{3}}%
\offprints{}          % Insert a name or remove this line
\institute{Dipartimento di Fisica, Universit\`a di Firenze and INFN, I-50019 Sesto F.no (FI), Italy \and 
           Dipartimento di Sistemi e Informatica and INFN, Universit\`a Di Firenze, Via S. Marta 3, I-50139 Florence, Italy  \and
           Centre de Biophysique Mol\'eculaire (CBM), CNRS-UPR 4301 and Universit\'e d'Orl\'eans, D\'epartement de Physique, \\
           45071 Orl\'eans cedex, France, \email{Francesco.Piazza@cnrs-orleans.fr} }
\date{Received:  }
\abstract{
Modelling the propagation of a pulse in a dense {\em milieu} poses 
fundamental challenges at the theoretical and applied levels. 
To this aim, in this paper we generalize the telegraph equation to non-ideal conditions by 
extending the concept of persistent random walk to account for spatial exclusion effects. 
This is achieved by introducing an explicit constraint in the hopping rates, that weights the occupancy of the target sites. 
We derive the mean-field equations, which display nonlinear terms that are important 
at high density. We compute the evolution of the mean square displacement (MSD)
for pulses belonging to a specific class of spatially symmetric initial conditions. 
The MSD still displays a transition from ballistic to diffusive behaviour. 
We derive an analytical formula for the effective velocity of the ballistic stage,
which  is shown to depend in a nontrivial fashion upon both the {\em density}
(area) and the {\em shape} of the initial pulse. 
After a density-dependent crossover time, nonlinear terms become negligible and 
normal diffusive behaviour is recovered at long times. 
\PACS{
      {05.60.Cd}{Classical transport}   \and
      {02.50.Ey}{Stochastic processes}
     } % end of PACS codes
} %end of abstract
\maketitle
%%%%%%%%%%%%%%%%%%%%%%%%%%%%%%%%%%%%%%%%%%%%%%%%%%%%%%%%%%%%%%%%%%%%%%%%%%%%%%%%%%%%%%%%%%%%%%%%%%%%%%%%%%%%%%%%%%%%%%%%%%%%%%%%%%%%%%%
\section{\label{s:0}Introduction}
%%%%%%%%%%%%%%%%%%%%%%%%%%%%%%%%%%%%%%%%%%%%%%%%%%%%%%%%%%%%%%%%%%%%%%%%%%%%%%%%%%%%%%%%%%%%%%%%%%%%%%%%%%%%%%%%%%%%%%%%%%%%%%%%%%%%%%%

Random walks (RW) are widely studied stochastic processes with a wealth of different
applications in many fields.~\cite{Haus:1987uq,Weiss:1994fk}. The persistent random walk (PRW) identifies
a special class of RWs, where agents also have a bias to keep hopping in the same direction as they did 
in the past (with finite memory). While the continuum limit of standard RWs is the diffusion equation, 
yielding infinite propagation velocity, the continuum limit of the PRW is the so-called telegraph equation,
which displays a transition from ballistic to diffusive transport at a characteristic time.  
The telegraph equation is also obtained from the so-called dichotomous (or sometimes also called {\em telegraph}) 
noise (random switches between two states) when the distribution of switching times is exponential 
and the process is drift-less~\cite{Gardiner:1983uq}.
\\
\indent The telegraph equation was first studied by Lord Kelvin,
who was interested in the distortion and dissipation
of electromagnetic waves in telegraph lines, motivated by 
the design of the first transatlantic cable~\cite{Masoliver:1996vn}, while
PRW and its connection with the telegraph equation were first studied 
by Goldstein in 1951~\cite{GOLDSTEIN:1951kx}. 
The telegraph equation is notoriously used in many contexts, 
from transport of relativistic particles~\cite{Dunkel:2007rp,Landau:1987kx} in different {\em milieux},
such as in multiply scattering media~\cite{PhysRevE.59.6517}, to second sound in liquid helium II 
and inertial effects in heat transport~\cite{RevModPhys.61.41,RevModPhys.62.375}.
An interesting discussion of many applications 
of the telegraph equation can be found in a recent review by G. Weiss~\cite{Weiss:2002vn}.
\\
\indent Macroscopic transport equations can be derived in a straightforward manner from {\em microscopic} stochastic
processes, the diffusion equation being the classical text-book example. In one dimension, for example, 
if $P_{i}(n)$ denotes the probability that an agent is at site $i$ on some discrete manifold at time 
$n\Delta t$, a simple unbiased RW corresponds to the update rule
\begin{equation}
\label{e:diffRW}
P_{i}(n) = \frac{1}{2} [P_{i-1}(n-1) + P_{i+1}(n-1)]
\end{equation}
as it is assumed that at each time step the walker can either jump to its right or to its left
with equal probability. Letting the lattice spacing $a$ and the time step $\Delta t$ go to zero,
such that $\lim_{a,\Delta t \to 0} (a^{2}/\Delta t) = D$, one obtains the diffusion equation
$\partial P(x,t)/\partial t = D \, \partial^{2}P(x,t)/\partial x^{2}$.
Of course, in performing the continuum limit one is tacitly assuming that many walkers are performing 
as many uncorrelated random walks and that a probability of being at $x$ at time $t$ can be defined
by averaging over such uncorrelated trajectories. This requires the walkers to be transparent to each other.
It is then interesting to ask the following question. If some exclusion rule is enforced, such that a walker 
can only jump on an {\em empty} site, how will the macroscopic equations be modified? And what kind
of process will they describe? The macroscopic counterpart of this problem implies 
transport in a non-ideal {\em milieu}, {\em e.g.} a dense fluid or a complex scattering medium.
Examples from optics include pulse diffusion in whole blood
and in a dense distribution of particulate matter in the atmosphere and the ocean.
Indeed it has been shown that the telegraph equation can 
can be derived as an approximation to a physically more realistic transport equation~\cite{Ishimaru:1978gd}.
\\
\indent Microscopic jump processes that implement exclusion rules go under the name of simple exclusion processes (SEP).
In general, SEPs are space-discrete, agent-based stochastic processes modeling  
some kind of transport according to specific rules and bound to the constraint 
that no two agents can ever occupy the same site. SEPs play a central role in non-equilibrium 
statistical physics~\cite{Privman:1997kl,Liggett:1999fk}. While the first theoretical ideas underlying 
such processes can be traced back to Boltzmann's works~\cite{Boltzmann:1966fk}, SEPs were introduced and 
widely studied in the 70s as simplified models of one-dimensional transport for phenomena like hopping
conductivity~\cite{Richards:1977tg} and kinetics of biopolymerization [5]. Along the same lines, 
the asymmetric exclusion process (ASEP), originally introduced by Spitzer~\cite{Spitzer:1970qo},
and the associated macroscopic mean-field equations~\cite{Simpson:2009ys} have 
become a paradigm in non-equilibrium statistical physics~\cite{Derrida:1993bh,Schutz:1993ye,Derrida:1998oq}
and has now found many applications, {\em e.g.} to the study of molecular motors~\cite{Golubeva:2012qf},
transport through nano-channels~\cite{Zilman:2010cr} and dynamics of microtubule
depolymerization~\cite{Reese:2011nx}.
\\
\indent Starting from this setting, we shall here generalize the concept of persistent random walk to  cases of interest 
where exclusion effects are to be accounted for. More precisely, we will introduce a modified PRW 
featuring an explicit constraint in the hopping probabilities, that are now gauged by 
the occupancy of the target sites. We will proceed on to deriving the mean-field equations for the 
concentration. These will be shown to display nonlinear terms, that prove however 
negligible in the diluted limit. Working at high densities, excluded-volume corrections 
do matter, as we shall here substantiate both analytically and numerically.
\\
\indent The paper is organized as follows. In Section~\ref{s:1} we  introduce the persistent simple exclusion process (PSEP) 
and derive the mean-field equations for the continuum densities. In Section~\ref{s:2} we  characterize the evolution 
of the  mean square displacement for a pulse belonging to a specific class of spatially symmetric initial conditions. 
Interestingly, as a result of the excluded-volume constraint, the ballistic-to-diffusive transition becomes dependent 
on the density.
Moreover, the effective velocity that characterizes the ballistic stage becomes a function of the crowding level too, 
but also turns out to depend on the {\em shape} of the initial pulse. 
Finally, in Section~\ref{s:3} we summarize our results and sketch possible interesting directions along which 
to pursue this work. 

%%%%%%%%%%%%%%%%%%%%%%%%%%%%%%%%%%%%%%%%%%%%%%%%%%%%%%%%%%%%%%%%%%%%%%%%%%%%%%%%%%%%%%%%%%%%%%%%%%%%%%%%%%%%%%%%%%%%%%%%%%%%%%%
\section{\label{s:1}The persistent simple exclusion process (PSEP)}
%%%%%%%%%%%%%%%%%%%%%%%%%%%%%%%%%%%%%%%%%%%%%%%%%%%%%%%%%%%%%%%%%%%%%%%%%%%%%%%%%%%%%%%%%%%%%%%%%%%%%%%%%%%%%%%%%%%%%%%%%%%%%%%

Let us consider a bunch of $N$ walkers on a one-dimensional lattice with spacing $d$ and length $L$. 
According to the definition of persistent random walk~\cite{GOLDSTEIN:1951kx},
at regular intervals $\Delta t$ a walker can jump in the same direction as it did at the previous step with 
probability $p$ or invert its direction with probability $q$. We take $q= 1-p$, which amounts to assuming that 
there is no leakage~\cite{GOLDSTEIN:1951kx} in the system. 
Let us denote with $a_i(n)$ the probability that a walker is at site 
$i$ at time $n \Delta t$ having been at site $i-1$ at time $(n-1)\Delta t$ (right-bound flow) and with 
$b_i(n)$ the probability that a walker is at site 
$i$ at time $n \Delta t$ having been at site $i+1$ at time $(n-1)\Delta t$ (left-bound flow). 
If walkers are invisible to each other, the following relations hold
\begin{eqnarray}
a_{i}(n) &= p \, a_{i-1}(n-1) + q\, b_{i-1}(n-1) \label{e:abstandard1}\\
b_i(n) &= p \, b_{i+1}(n-1) + q\, a_{i+1}(n-1) \label{e:abstandard2}
\end{eqnarray}
The above equations describe a discrete stochastic process. The continuum limit can be obtained by introducing the 
continuous probability density $P(x,t) = \langle P_{i}(n) \rangle \equiv 
\langle a_i(n)+b_i(n)\rangle$, where $\langle \dots \rangle$ denotes an
average over the trajectories of many agents, by letting $d\to0$,$\Delta t\to0$,$q\to0$. By doing this, 
it is known that one gets the telegraph equation~\cite{Weiss:2002vn}
\begin{equation}
\label{e:telegraph}
\frac{\partial^{2} P}{\partial t^{2}} + 2r \, \frac{\partial P}{\partial t} = v^{2} \frac{\partial P}{\partial x}  
\end{equation}
with 
\begin{equation}
\label{e:vqlim}
\lim_{d,\Delta t\to0} \frac{p}{\Delta t} = v \qquad 
\lim_{q,\Delta t\to0} \frac{q}{\Delta t} = r 
\end{equation}
In this paper we wish to study how the persistent random walk is modified if we introduce the constraint that no two walkers 
can occupy the same site at the same time. That is, the probability to jump to a given site is gauged 
by the current occupancy of that site. Along the same line of reasoning of SEPs and ASEPs, we
modify Eqs.~\eref{e:abstandard1} and~\eref{e:abstandard2} in the following way
\begin{eqnarray}
a_i(n) - a_{i}(n-1) &=&   [p \, a_{i-1}(n-1) +   q\, b_{i-1}(n-1)] \times \nonumber\\&&[1-P_{i}(n)]
                          - a_{i}(n-1) \times \nonumber \\ 
                         && \{p [ 1 - P_{i+1}(n-1)]  + \label{e:abmod1} \\ && \ \ \ \ q [ 1 - P_{i-1}(n-1)]\} \nonumber \\
b_i(n) - b_{i}(n-1) &=&    [p \, b_{i+1}(n-1) + q\, a_{i+1}(n-1)]\times \nonumber\\ && [1-P_{i}(n)] 
                           -b_{i}(n-1) \times \nonumber\\ &&
                            \{p [ 1 - P_{i-1}(n-1)]  + \label{e:abmod2}\\ && \ \ \ \  q [ 1 - P_{i+1}(n-1)]\} \nonumber
\end{eqnarray}
where $P_{i}(n) = a_{i}(n) + b_{i}(n)$. Again, the idea is to gauge jump probabilities by the occupancy of the
target sites. For example, the first term in the r.h.s of eq.~\eref{e:abmod1} states that a net increase of the probability at site $i$
associated with the right-bound flow is only possible with a transition rate proportional to the amount of
{\em free room} at site $i$, {\em i.e.} $1 - P_{i}$. If all walkers happen to be at site $i$ at the same time, then 
$P_{i}=1$ and no further increase of $a_{i}$ nor of $b_{i}$ is possible. 
In order to take the continuum limit,  we first 
divide Eqs.~\eref{e:abmod1} and~\eref{e:abmod2} by $\Delta t$ and substitute 
$q = 1 - p$. Then, recalling the definitions~\eref{e:vqlim}, we get
\begin{eqnarray}
\label{e:telegrc1}
\frac{\partial a}{\partial t} + v \frac{\partial }{\partial x} [a(1-P)] &=& -r J(1-P) \nonumber\\
\frac{\partial b}{\partial t} - v \frac{\partial }{\partial x} [b(1-P)] &=&  r J(1-P) 
\end{eqnarray}
where $P(x,t) = a(x,t) + b(x,t)$ and $J(x,t) = a(x,t) - b(x,t)$.
\\
\indent For the sake of the argument, let us consider the propagation of pulses in a fluid,
{\em i.e.} travelling  density fluctuations. 
Eqs.~\eref{e:telegrc1} contain the single-particle probability field $P$, which is a number between zero
and one. The value $P=1$  should then correspond to the maximum density allowed in the system.
Thus,  more {\em physical} equations can be obtained by introducing the agent densities
\begin{eqnarray}
\rho(x,t) \equiv \rho_{\scs M} P(x,t) \label{e:rho}\\
\mathcal{J}(x,t) \equiv \rho_{\scs M} J(x,t) \label{e:J}\\
\rho_{+}(x,t) = \rho_{\scs M} a(x,t) \label{e:a}\\
\rho_{-}(x,t) = \rho_{\scs M} b(x,t) \label{e:b}  
\end{eqnarray}
where $\rho_{\scs M}$ is the maximum allowed density, which in principle could be regarded 
as a parameter of the model. If we imagine that the  agents have a finite size $\sigma$, {\em i.e.} we 
regard them as hard rods~\footnote{It should be emphasized that it is in principle possible to 
write mean-field equations for an exclusion process that accounts for {\em extended} objects on a line 
from the beginning. In ref.~\cite{Schonherr:2004zr} the authors derive a modified diffusion-advection
equation from a microscopic exclusion process (RW with drift) for hard rods. 
However, even if it would be intriguing to do so, 
it is rather unclear how to apply the same technique in the context of the PRW.}, 
one simply have $\rho_{\scs M} = 1/\sigma$. 
%
%The density $\rho$ has dimensions of an inverse length and it is normalized so that
%
%\begin{equation}
%\label{e:normrho}
%\lim_{N,L\to\infty}\frac{1}{L}\int_{0}^{L}\rho(x,t)\, dx = \lim_{N,L\to\infty}\frac{N}{L} \equiv %\rho_{0} 
%\end{equation}
%
%where $\rho_{0}$ is the bulk density of agents.
Introducing the density $\rho_{\scs M}$, eqs.~\eref{e:telegrc1}  become
\begin{eqnarray}
\label{e:telegrc2}
\frac{\partial \rho_{+}}{\partial t} + v \frac{\partial }{\partial x} 
                                      \left[\rho_{+}\left(
                                          1-\frac{\rho}{\rho_{\scs M}}
                                          \right)\right] = -r \mathcal{J} \left(
                                          1-\frac{\rho}{\rho_{\scs M}}
                                          \right)\nonumber\\
\frac{\partial \rho_{-}}{\partial t} - v \frac{\partial }{\partial x} 
                                         \left[\rho_{-}\left(
                                          1-\frac{\rho}{\rho_{\scs M}}
                                          \right)\right] = r \mathcal{J} \left(
                                          1-\frac{\rho}{\rho_{\scs M}}
                                          \right) 
\end{eqnarray}
A system of equations for the densities  $\rho(x,t)$ and $\mathcal{J}(x,t)$
can be obtained by adding and subtracting the two equations~\eref{e:telegrc2}
\begin{eqnarray}
\label{pippo}
\frac{\partial\rho}{\partial t}+v \frac{\partial}{\partial x}
                       \left[
                             \mathcal{J}\left(1-\frac{\rho}{\rho_M}
                                        \right)
                       \right]=0 
\nonumber\\
\frac{\partial \mathcal{J}}{\partial t}+v \frac{\partial}{\partial x}
                       \left[
                              \rho\left(1-\frac{\rho}{\rho_M}
                                  \right)
                       \right] = -2r\mathcal{J}\left
                                               (1-\frac{\rho}{\rho_M}
                                                \right)
\end{eqnarray}
As a general remark, we see that the microscopic exclusion constraint results in the appearance 
of nonlinear terms. The standard evolution of the PRW leading to the 
telegraph equation is obtained in the {\em dilute} limit $\rho \ll \rho_M$.
Conversely, we may consider the full system~\eref{pippo} as describing 
transport in a {\em crowded} medium. More precisely, the nonlinear equations~\eref{pippo}
embody the microscopic excluded-volume constraint that emerges at
high densities.

%%%%%%%%%%%%%%%%%%%%%%%%%%%%%%%%%%%%%%%%%%%%%%%%%%%%%%%%%%%%%%%%%%%%%%%%%%%%%%%%%%%%%%%%%%%%%%%%%%%%
\section{\label{s:2} Mean square displacement}
%%%%%%%%%%%%%%%%%%%%%%%%%%%%%%%%%%%%%%%%%%%%%%%%%%%%%%%%%%%%%%%%%%%%%%%%%%%%%%%%%%%%%%%%%%%%%%%%%%%%

We turn now to analyzing how the excluded-volume constraint affects the propagation 
in an infinite medium of an initially localized pulse.
It is well known that the  PRW displays a transition from ballistic to diffusive 
transport, as exemplified by the mean square displacement (MSD),
\begin{equation}
\label{e:MSD}
\mu_2(t)\equiv\frac{1}{\mathcal{N}}  
[\langle x^2(t) \rangle_{\rho} - \langle x(t) \rangle^{2}_{\rho} ] 
\end{equation} 
with $\langle x^m(t) \rangle_{\rho} = \int x^m \rho(x,t)\,dx$ and 
$\mathcal{N} = \int \rho(x,t)\,dx$.
As it is customarily done, we shall here restrict to a class of symmetric initial pulses, 
namely such that $\rho(x,t=0)=\rho(-x,t=0)$,
$\mathcal{J}(x,t=0)=0$, meaning that the initial distribution of the right-headed 
agents is equal to that of the left-headed ones. In this case, it is straightforward to  
show that $\langle x (t) \rangle \equiv 0$ $\forall \, t$.

For the PRW  one has
\begin{eqnarray}
\mu_{2}(t) - \mu_{2}(0) &=& \frac{v^2}{2r^2}(2rt-1+e^{-2rt}) \nonumber\\
&\simeq &
\left\{
\begin{array}{ll}
v^2t^2 &\mbox{\rm for} \quad t\ll 1/2r
\\
\left(\frac{\displaystyle v^2}{\displaystyle r}\right) t &\mbox{\rm for} \quad t\gg 1/2r
\end{array}
\right.
\label{e:MSDdilute}
\end{eqnarray}
When excluded-volume effects are important, it appears impossible 
to obtain a closed expression for $\mu_2(t)$.
However, one can still capture the asymptotic regime at short times for a certain class of 
symmetric initial conditions in an infinite medium  (see ref.~\cite{Masoliver:1996vn}
for a critical discussion of reflecting and absorbing 
boundary conditions for the telegraph equation in a bounded domain).
Let us consider the Taylor expansion of $\mu_2(t)$
\begin{equation}
\mu_2(t)=\mu_2(0) + \mu_2^{\,\prime}(0)\,t+ 
\frac{1}{2}\mu_2^{\,\prime\prime}(0)\,t^2 + O(t^3)
\end{equation}
In order to evaluate the coefficients of the expansion, 
let us multiply the first equation of eq.~\eref{pippo} 
by $x^2$ and the second one by $x$ and integrate. 
Integrating by parts and assuming that boundary terms vanish, we obtain
\begin{eqnarray}
\label{e:sistemamomenti}
\frac{\ds d}{\ds d t }\langle x^2\rangle_{\rho} - 2v \langle x \rangle_{\mathcal{J}} +
\frac{2v}{\rho_M} \langle x \rangle_{\mathcal{J}\rho}=0   \nonumber\\
\frac{\ds d} {\ds d t} \langle x \rangle_\mathcal{J} -
v \int \rho\left(1-\frac{\rho}{\rho_M}\right)dx + \nonumber\\2r \langle x \rangle_\mathcal{J}-
\frac{2r}{\rho_M}\langle x\rangle_{\mathcal{J}\rho}=0
\end{eqnarray}
where $\langle \dots \rangle_\mathcal{J}$ and $\langle \dots \rangle_\mathcal{J\rho}$
denote averages with respect to the corresponding (products of) densities.
\\
\indent Since the initial condition is symmetric, the first equation
of~\eref{e:sistemamomenti} shows that $\mu_2^{\,\prime}(0)=0$.
Differentiating the same equation with respect to time,
recalling equations~\eref{pippo} and integrating by parts eventually leads to 
\begin{eqnarray}
\frac{d^{2}}{dt^{2}} \langle x^{2}\rangle_{\rho}= 
2v \frac{d}{dt}\langle x\rangle_\mathcal{J}-\frac{2v}{\rho_M}\frac{d}{dt}\langle x\rangle_{\mathcal{J}\rho}  \ \ \ \ \ \ \ \ \ \ \ \nonumber\\
=2v\bigg(v\int\rho\bigg(1-\frac{\rho}{\rho_M}\bigg)dx- \ \ \ \ \ \ \ \ \ \ \ \   \nonumber\\ 2r\langle x\rangle_\mathcal{J}+\frac{2r}{\rho_M}\langle x \rangle_{\mathcal{J}\rho}\bigg) - \ \ \ \ \ \ \  \nonumber\\
\frac{2v}{\rho_M}\bigg(
v\int \rho^2\bigg(1-\frac{\rho}{\rho_M}\bigg) dx+ \ \ \ \ \ \nonumber\\ 
v\int x \rho \bigg(1-\frac{\rho}{\rho_M}\bigg)\frac{\partial\rho}{\partial x} dx \bigg)
\end{eqnarray}
Evaluating the previous expression at $t=0$ makes the terms that 
involve the function $\mathcal{J}$ disappear. Thus
\begin{eqnarray}
\left. \frac{d^{2}}{dt^{2}} \langle x^{2}\rangle_{\rho} \right|_{t=0}
= 2v^2\int\rho\bigg(1-\frac{\rho}{\rho_M}\bigg)^2dx\bigg|_{t=0}- \ \ \ \ \ \nonumber\\
\frac{2v^2}{\rho_M}\int x \rho \bigg(1-\frac{\rho}{\rho_M}\bigg)\frac{\partial\rho}{\partial x} dx\bigg|_{t=0}
\end{eqnarray}
which leads to the approximation
\begin{equation}
\label{e:MSDapprox}
\mu_2(t) \approx \mu_2(0) + v_{\rm e}^2\,t^2
\end{equation}
with
\begin{eqnarray}
\label{e:veff}
v_{\rm e} =  \frac{v}{\sqrt{\mathcal{N}}}\left[
           \int \rho\bigg(1-\frac{\rho}{\rho_M}\bigg)^2 dx\bigg|_{t=0}- \right. \ \ \ \ \ \ \ \ \ \ \ \ \ \nonumber\\ \left. 
           \frac{1}{\rho_M}\int x
           \rho\bigg(1-\frac{\rho}{\rho_M}\bigg)\frac{\partial \rho}{\partial x } \,dx \bigg|_{t=0}
           \right]^{1/2}
\end{eqnarray}
where we have used the fact the norm 
$\mathcal{N} = \int \rho(x,t)\,dx$ is constant if boundary terms can be neglected.
This is the case of a broad choice of initial conditions, such as the propagation 
of pulses that are initially localized in a compact domain.
%
%=====================================================================================================
\begin{figure}[t!]
\centering
\includegraphics[width=\columnwidth]{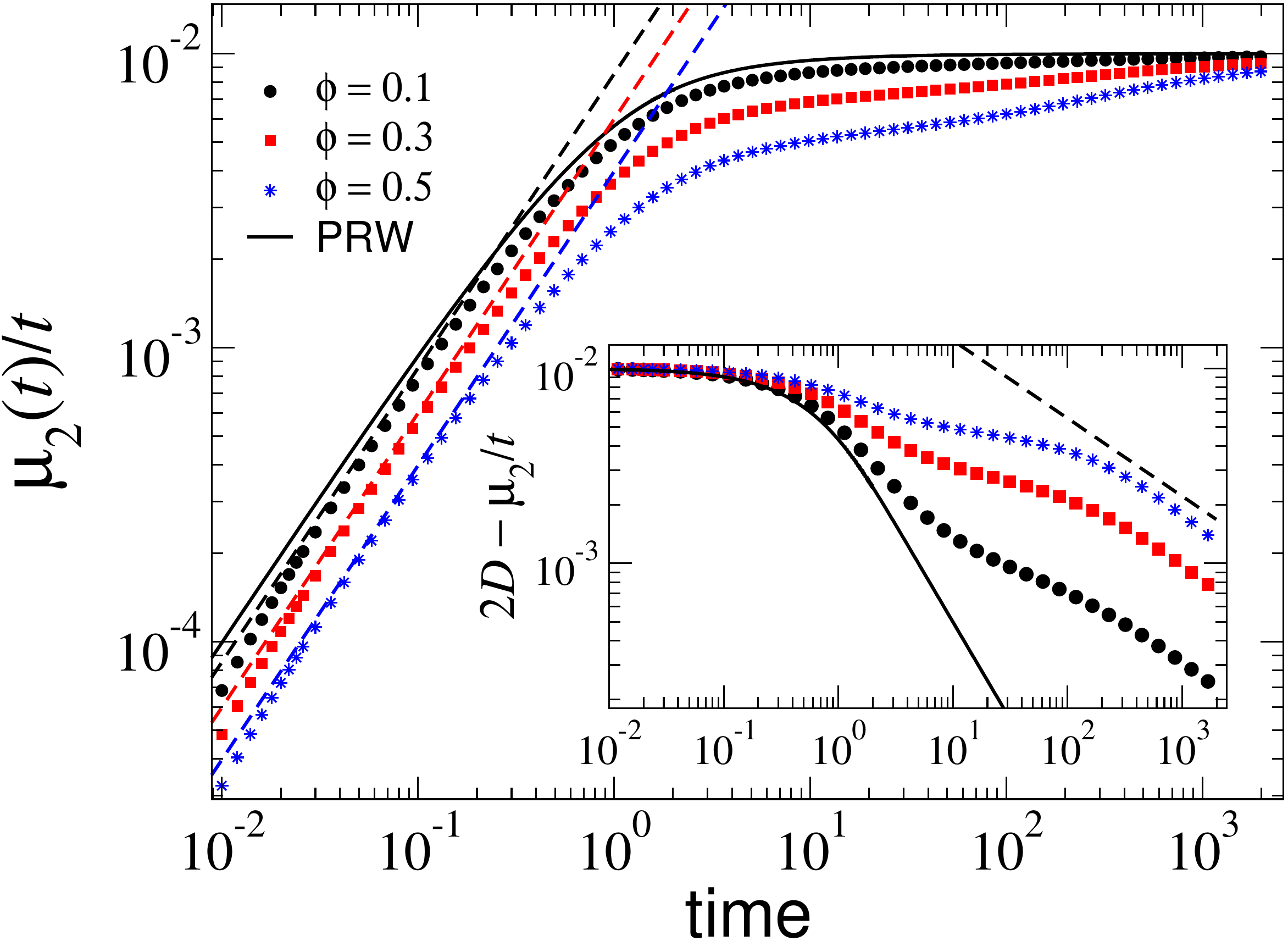}
\caption{Mean square displacement for a pulse of the kind~\eref{e:pulse} with $\beta=10$
as a function of time for different choices of the crowding parameter 
$\phi$. The dashed lines are straight lines of slope $v_{\rm e}$ as predicted by eq.~\eref{e:vphi}.
Inset: approach to the diffusive regime. The parameter $D=v^{2}/2r$ is the theoretical 
diffusion coefficient. The solid line refers to the PRW and vanishes as $t^{-1}$.
The dashed line is an inverse-power law with exponent $0.4$, to be used as a guide for the eye.}
\label{f:MSD}
\end{figure}
%=====================================================================================================
%=====================================================================================================
\begin{figure}[t!]
\centering
\includegraphics[width=\columnwidth]{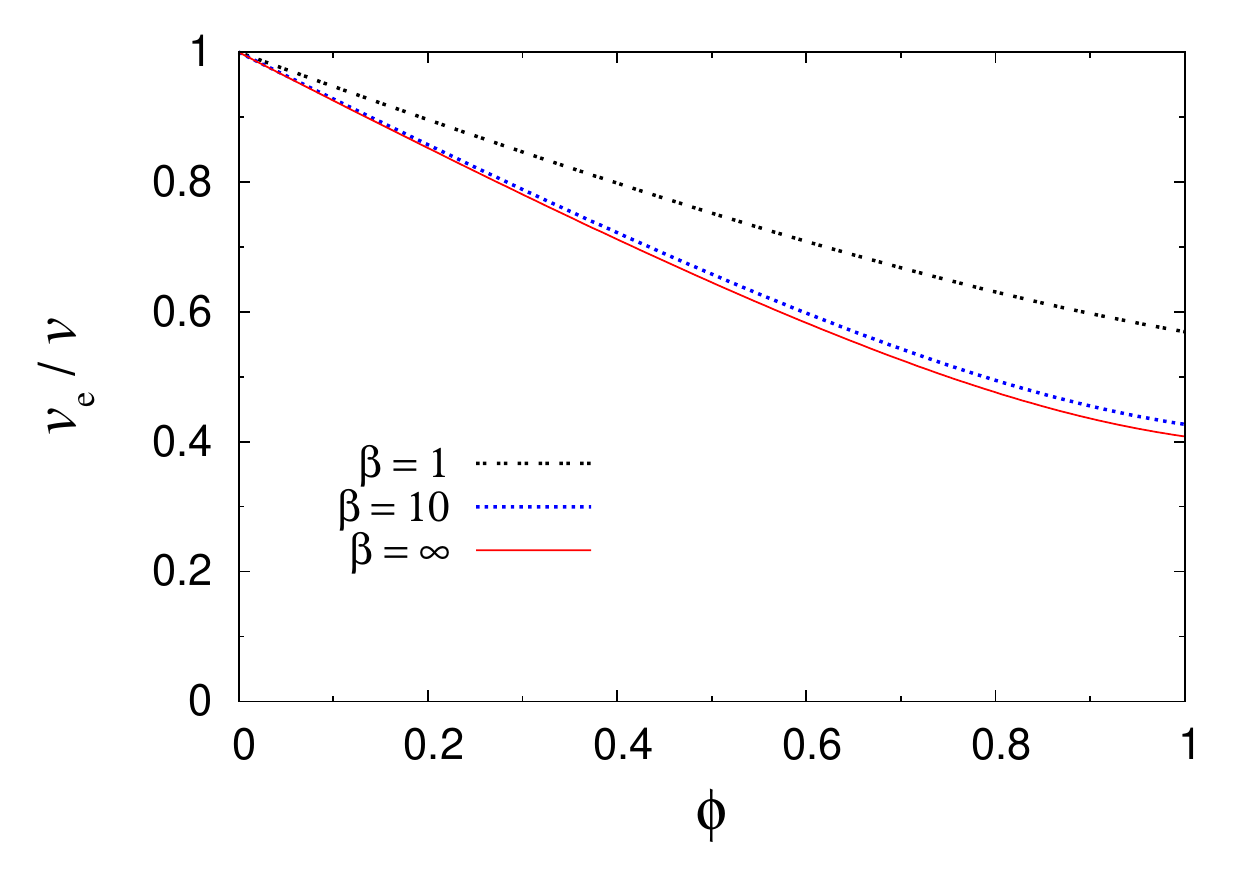}
\caption{Effective velocity~\eref{e:vphi} as a function of the level of crowding for a Gaussian 
pulse, a generalized Gaussian with $\beta = 10$ and a step pulse ($\beta \to \infty$).}
\label{f:vphi}
\end{figure}
%=====================================================================================================
\\
\indent Fig.~\ref{f:MSD} shows the time evolution of $\mu_{2}$ obtained by integrating numerically 
the system~\eref{e:sistemamomenti} with a forward-difference approximation in time 
and replacing the spatial derivatives by centered Euler approximations.
The initial conditions are generalized Gaussian pulses of the type
\begin{equation}
\label{e:pulse}
\rho(x,t=0) = \phi \, e^{-x^{2\beta}/2}
\end{equation}
This allows us to investigate the propagation of a pulse whose shape varies 
continuously from Gaussian  ($\beta=1$) to a sharp, nearly piecewise constant 
step function  ($\beta\gg 1$), while the parameter $\phi<1$ gauges the level of crowding.
It is not difficult to compute $v_{\rm e}$ analytically from eq.~\eref{e:veff} 
as a function of $\phi$ for a pulse of the kind~\eref{e:pulse}. After straightforward 
calculations, one gets
\begin{equation}
\label{e:vphi}
v_{\rm e}(\phi) = v \left[
                         1 - \frac{3 \, \phi}{2^{\gamma}} + 
                             \frac{2 \, \phi^{2}}{3^\gamma} 
                    \right]^{1/2}
\end{equation}
where $\gamma = 1+1/2\beta$. It is clear from the figure that the approximation~\eref{e:vphi} captures 
to an excellent extent the initial ballistic stage. Furthermore, the numerical 
integration of eqs.~\eref{e:sistemamomenti} clearly shows that asymptotically the propagation 
becomes diffusive, with the same diffusion coefficient $v^{2}/2r$ as the PRW. 
This is to be expected as $\rho\to 0$ as $t\to\infty$ and therefore the excluded-volume
constraints (that is, the nonlinear terms) become negligible. Nevertheless, the inset 
in fig.~\ref{f:MSD} clearly shows that the approach to the diffusive regime 
is considerably slowed down as a result of crowding -- the more the greater the 
excluded-volume constraint. 
\\
\indent Our analysis shows that  the initial behaviour of the mean square displacement 
is qualitatively the same as in the PRW, {\em i.e.} the propagation is ballistic.
The effect of crowding is to decrease the velocity that characterizes the initial stage of the evolution.  
In Fig. \ref{f:vphi} the effective velocity $v_{\rm e}$, normalized to the diluted limit $v$, is plotted as function of the
level of crowding and for different choices of the parameter $\beta$. Remarkably, $v_{\rm e}$ also depends on the {\em shape} of the 
initially localized density pulse. The ballistic spreading of a super-Gaussian pulse, nearly a 
sharp step, proceeds with a considerably lower speed as compared to the spreading 
of a pure Gaussian pulse (see again eq.~\eref{e:veff}).

%===============================================================================================
\section{Conclusions and perspectives}
\label{s:3}

The persistent random walk yields the so-called telegraph equation in the continuum limit, which
displays a well known transition from ballistic to diffusive transport. In the classical microscopic formulation 
of the PRW, individual walkers are assumed to jump toward neighboring sites with constant probability. 
However, one can account for excluded volume effects by gauging the hopping 
probability  to the occupancy  of the target sites. 
By doing so, we obtained a set of two coupled nonlinear transport equations, 
that reflect the microscopic competition for the available space. 
Microscopic processes which implement exclusion rules are called
simple exclusion processes (SEPs). For this reason, we refer to the generalized model  introduced here as  the 
persistent simple exclusion process (PSEP). It is interesting to remind here a rather surprising fact.
From the above discussion, one may think that a modified diffusion equation including finite-density effects could be derived 
through occupancy-gauged hopping rules such as those appearing in eqs.~\eref{e:abmod1} and~\eref{e:abmod2}
from a standard RW process without drift (eq.~\eref{e:diffRW}). However, as first noticed by Richards in 1977~\cite{Richards:1977tg},
nonlinear terms cancel exactly in doing this, and one is left with the standard diffusion equation in the continuum limit.
\\
\indent We go on to investigate the mean-field limit of the PSEP process for a specific class of initial conditions. These are 
generalized Gaussian pulses, whose shape varies continuously from Gaussian to sharp steps depending on a 
control parameter. The pulse amplitude $\phi$ in this setting measures the degree of imposed crowding,
{\em i.e.} the density of the medium.
Numerical integration of our modified transport equations shows that the PSEP 
still undergoes a transition from ballistic to diffusive behaviour of the MSD. 
However, the velocity of the initial ballistic stage is found to decrease with the density of the medium ($\phi$).
This might be relevant in many cases where the the telegraph equation is used to model physical situations. 
Think for example to gel electrophoresis (GEP), first modeled as a two-state process yielding the telegraph 
equation in the 50s~\cite{Giddings:1955fk}. 
As a molecules moves along a channel in response to an applied electric field,
it may become entangled in the gel matrix at random times and after some time detach from 
the gel fibers because of thermal fluctuations. Here $a(x,t)$ and $b(x,t)$ represent the 
concentration of bound and free molecules. Of course, the available theory does not describe the migration 
of molecules in crowded solutions, which may be however interesting to analyze through GEP.
\\
\indent Remarkably, as a consequence of the excluded-volume constraint,
the velocity of the ballistic stage also depends on the {\em shape} of the initial pulse.
In particular, at equal values of crowding $\phi$, our calculations show that 
pulses with sharper edges display smaller velocities and hence cross over later to diffusive spreading. 
At long times, the propagation becomes diffusive, with the same diffusion coefficient as for the long-time limit of the 
telegraph equation. This is because, as the pulse spreads, the density decreases and nonlinear 
terms become eventually irrelevant. 
\\
\indent We envisage to extend this work along different lines. On the one side it would be interesting to account 
from the very beginning for the finite size of the agents along the lines of~\cite{Schonherr:2004zr}, 
beyond the point-like version of the crowding considered here ({\em i.e.} fully penetrable entities).
Moreover, it would also be engaging to establish a link between our treatment and 
the dichotomous noise picture, in order to investigate the  connection between excluded-volume 
constraints and modifications of the switching time statistics.

%%%%%%%%%%%%%%%%%%%%%%%%%%%%%%%%%%%%%%%%%%%%%%%%%%%%%%%%%%%%%%%%%%%%%%%%%%%%%%%%%%%%%%%%%%%%%%%%
%  BIBLIOGRAPHY
%
\bibliographystyle{unsrt}
\bibliography{crowding,bio-11} 
%%%%%%%%%%%%%%%%%%%%%%%%%%%%%%%%%%%%%%%%%%%%%%%%%%%%%%%%%%%%%%%%%%%%%%%%%%%%%%%%%%%%%%%%%%%%%%%%%

%
% BibTeX users please use
% \bibliographystyle{}
% \bibliography{}
%

\end{document}